\documentclass[aps,prx,superscriptaddress,twocolumn]{revtex4-1}
\pdfoutput=1
\usepackage{amsmath}
\usepackage{amssymb}
\usepackage{graphicx}
\usepackage{xcolor}
\usepackage{braket}
\usepackage{verbatim}
\usepackage{marginnote}

\begin{document}

\newcommand{\beq}{\begin{equation}}
\newcommand{\eeq}{\end{equation}}
\newcommand{\beqa}{\begin{eqnarray}}
\newcommand{\eeqa}{\end{eqnarray}}
\newcommand{\ben}{\begin{enumerate}}
\newcommand{\een}{\end{enumerate}}
\newcommand{\hs}{\hspace{0.5cm}}
\newcommand{\vs}{\vspace{0.5cm}}
\newcommand{\notemiles}[1]{{\color{blue}[Miles: #1]}}
\newcommand{\noteroger}[1]{{\color{red}[Roger: #1]}}
\newcommand{\notejm}[1]{{\color{violet}[JM: #1]}}
\newcommand{\tim}{$\times$}
\newcommand{\bigo}{\mathcal{O}}
\newcommand\leftnote[1]{\reversemarginpar\marginnote{\tiny \textcolor{blue}{#1}}}
\newcommand\rightnote[1]{\normalmarginpar\marginnote{\tiny \textcolor{blue}{#1}}}

\title{Unusual Corrections to Scaling and Convergence \\ of Universal Renyi Properties at Quantum Critical Points} 

\author{Sharmistha Sahoo}
\affiliation{Physics Department, University of Virginia, Charlottesville, VA 22904-4714, USA}

\author{E.\ Miles Stoudenmire}
\affiliation{Perimeter Institute for Theoretical Physics, Waterloo, Ontario, N2L 2Y5, Canada}

\author{Jean-Marie St\'ephan}
\affiliation{Max Planck Institute for the Physics of Complex Systems, N\"othnitzer Str.~38, 01187
  Dresden, Germany}

\author{Trithep Devakul}
\affiliation{Department of Physics, Princeton University, Princeton, NJ 08544, USA}

\author{Rajiv R.~P.~Singh}
\affiliation{Department of Physics, University of California Davis, CA 95616, USA}

\author{Roger G.\ Melko}
\affiliation{Perimeter Institute for Theoretical Physics, Waterloo, Ontario, N2L 2Y5, Canada}
\affiliation{Department of Physics and Astronomy, University of Waterloo, Ontario, N2L 3G1, Canada}

\date{\today}

\begin{abstract}
At a quantum critical point, bipartite entanglement entropies have universal quantities which are
subleading to the ubiquitous area law.  For Renyi entropies, these terms are known to be similar to the 
von Neumann entropy, while being much more amenable to  numerical and even experimental measurement.
We show here that when calculating universal properties of Renyi entropies, it is important to account for unusual corrections
to scaling that arise from relevant local operators present at the conical singularity in the multi-sheeted Riemann surface. These corrections 
grow in importance with increasing Renyi index. We present studies of Renyi correlation functions in the $1+1$ transverse-field Ising model
 (TFIM) using conformal field theory, mapping to free fermions, and series expansions, 
and the logarithmic entropy singularity at a corner in $2+1$ for both free bosonic field theory and the TFIM, using 
numerical linked
cluster expansions.  In all numerical studies, accurate results are only obtained when unusual corrections to scaling are taken into account.
In the worst case, an analysis ignoring these corrections can get qualitatively incorrect answers, such as predicting
a decrease in critical exponents with the Renyi index, when they are actually increasing. We discuss a two-step extrapolation
procedure that can be used to account for the unusual corrections to scaling.

\end{abstract}

\maketitle

\section{Introduction}

Quantum entanglement in the ground state of a many-body system contains universal signatures of a variety of low-energy and
long-lengthscale physical phenomena. These include Goldstone modes due to spontaneous breaking of a continuous symmetry \cite{Kallin:2011,Max_Tarun,Luitz_2015, Bohdan}; 
topological order \cite{Kitaev:2006,Levin:2006, Isakov_Topo}; Fermi surfaces \cite{Wolf_2006,Gioev_Klich}; quantum criticality \cite{Metlitski:2009,Casini:2012,Grover:2014e,Bueno:2015}; and more.
If a system is divided into two spatial regions $A$ and $B$,
the entanglement in a pure state
can be characterized by the von~Neumann entropy associated with the reduced density matrix of either subsystem,
\begin{equation}
S_{1} = - {\rm Tr} \rho_A \ln{\rho_A}.\label{vonN}
\end{equation}
However, numerically computing the von~Neumann entropy is a significant challenge, requiring an explicit representation of the reduced
density matrix or the ground state wavefunction.  
In continuum field theories, the von~Neumann entropy can be computed from the Renyi entropies,
\begin{equation}
S_{\alpha} = \frac{1}{1 - \alpha}\ln({\rm Tr} \rho_A^{\alpha}), 
\end{equation}
by introducing $\alpha$ identical replicas of the system, then
formally taking the limit $\alpha\to 1$ \cite{Calabrese:2004}.
Renyi entropies for integer values of $\alpha\ge 2$ can be
calculated by a variety of methods employing this replica-trick,
however, the limit $\alpha\to 1$ is difficult to obtain from numerical data for finite, discrete $\alpha$,
which can often be most accurate for small $\alpha$.

For this reason, integer $\alpha > 1$ Renyi entropies are often studied in quantum many-body systems on their own merit.
A variety of recent numerical techniques have been developed for this task.
One way to calculate them in Quantum Monte Carlo (QMC) simulations is by
evaluating the expectation value of a swap operator between the different replicas \cite{Hastings:2010}. 
Alternatively, these integer Renyi entropies can be expressed as a ratio of two
partition functions \cite{Melko:2010} and evaluated by stochastically sampling an extended ensemble where one switches between the configurations
defining the two partition functions \cite{Tommaso}. 
Such sampling techniques make the Renyi entropies available in other numerical approximations of the 
wave function, such as tensor network states \cite{Wang_2013}.	
They can also be calculated by series expansions in various coupling constants
of the model \cite{Singh:2011m,Kallin:2011,Singh:2012t,Devakul:2014,Devakul:2014e}. 
Most importantly, there are a number of proposals \cite{Cardy:2011,Abanin:2012,Daley_2012} to measure the Renyi entropy
experimentally.

This manuscript considers the universal properties associated with the Renyi entropies and associated Renyi
correlation functions in a critical system. While Renyi entropies do not
satisfy strong subadditivity {\it a priori}, and thus may not provide as rigorous an information measure of entanglement as Eq.~(\ref{vonN}),
explicit calculations have found that their singular behavior is very analogous to the von Neumann entropy in typical many-body systems.
Universal terms associated with Goldstone modes, topological order, Fermi surfaces and quantum critical points
often depend in a simple parametric way on the Renyi index $\alpha$. Hence, computing universal terms in any one Renyi entropy
can be sufficient to determine the universality class of the system. If there is an entanglement property that exhibits monotonicity under
renormalization \cite{Zamolodchikov:1986}, one would expect it to remain valid for the Renyi entropies as well.

However, there is a subtlety in obtaining universal, asymptotic Renyi properties, 
which is the main focus of this work.
Renyi entropies, and associated correlation functions, suffer from ``unusual'' corrections
to scaling that are absent for von Neumann entropies  \cite{CardyCalabrese2010}. These corrections become more and more severe as the Renyi index increases
and cause a slow convergence to the asymptotic limit. 
Rapid convergence in the von Neumann entropy and slower convergence for higher Renyi cases were
previously noted for topological  entanglement entropy, where they were found to be related to non-universal perturbative terms
\cite{Jiang:2013}.
Here, we discuss  a two step extrapolation process needed to obtain
the asymptotic universal critical Renyi properties. 

In light of our work, there may be a need to revisit previous calculations of Renyi entropies and their universal terms in
past computational studies.
For example, in this paper we re-examine recent numerical studies of $O(N)$ models in $2+1$ dimension, 
where it was found that the coefficients of the logarithmic singularities associated
with a corner approximately scales with $N$ \cite{Kallin:2013,Kallin:2014,Miles:2014,Helmes:2014}.
These studies also found significant differences between 
the entropy of $N$ free fields and a single copy of the interacting $O(N)$ theory (differences of order of greater than 10 percent). 
In light of a recent conjecture that the corner entanglement mimics a central charge $C_T$ \cite{Bueno:2015},
one might expect a much smaller difference.
Here, we find that some of the observed difference could be due to the unusual corrections
to scaling. Our work also suggests that one should be especially careful in computing universal properties
at large values of the Renyi index $\alpha$.

Following a discussion of how unusual corrections to Renyi properties arise in Section~\ref{UnusualSection}, we illustrate their importance
in Section~\ref{RenyiCorrSection} by studying Renyi correlation functions in (space+time) dimension $D=1+1$, for which there is a well-developed theory.
In Section~\ref{3Dcorn} we apply these insights in $D=2+1$ to universal entanglement entropy contributions from sharp
corners. Accounting for the unusually strong Renyi finite-size corrections significantly improves our numerical results for the universal
corner term for the free boson theory; we apply the same procedure to improve our estimates of the corner term for the 
critical Ising model.

\begin{figure}[t]
\includegraphics[width=0.55\columnwidth]{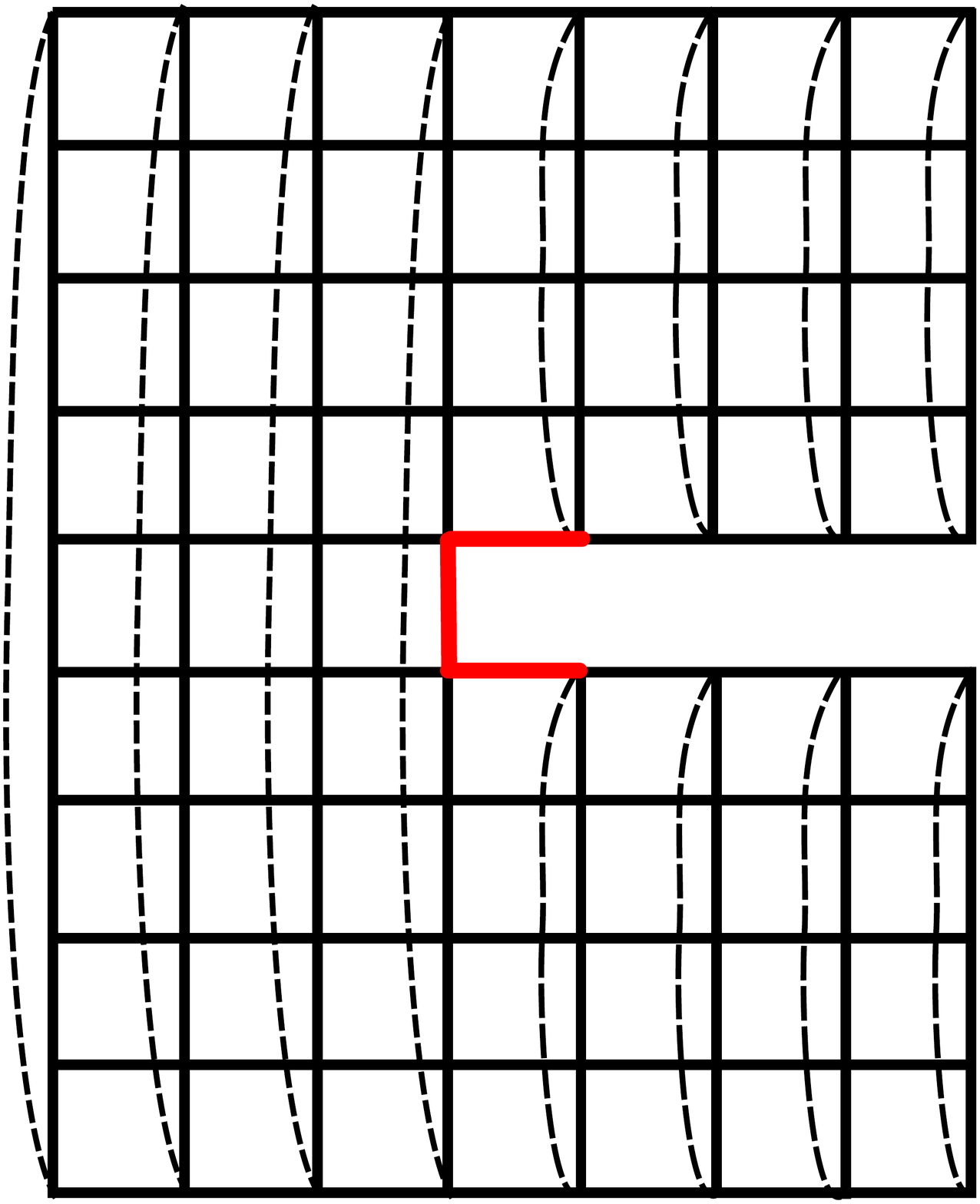}
\caption{A schematic view of the conical singularity for the $\alpha=2$ Renyi partition function $\mathcal{R}_2$.  Dashed lines represent the periodic boundaries in imaginary time.  All lattice sites retain a coordination number of four, however the re-arrangement of periodic boundaries in imaginary time near the conical
singularity (red) push that local region off-critical.}
\label{fig:conical_sing}
\end{figure}

\section{Unusual corrections to scaling} \label{UnusualSection}

Corrections to critical scaling are well known for bulk quantities and are due to the presence of irrelevant operators \cite{Wegner:1972}.
When approaching a critical point, a thermodynamic quantity such as the order parameter susceptibility $\chi$
does not behave as a single power-law, but as a sum of power laws. For small non-zero $t \propto T-T_c$,
\begin{equation}
\chi(t) = A t^{-\gamma} (1+ B t^{\theta^\prime}+\ldots), \label{CHI_t}
\end{equation}
or for a finite system of size $L$ at the critical point,
\begin{equation}
\chi(L) = A L^{\gamma^\prime} (1+ C /  L^{\theta}+\ldots) \:. \label{CHI_L}
\end{equation}
The subleading terms above lead to violations of simple scaling and data collapse, such as the failure of Binder ratios
to cross precisely at a critical point. 
The subleading terms can be seen to directly affect the calculation of exponents, by
 taking a logarithm of Eq.~(\ref{CHI_t}) and differentiating with respect to $t$ for small $t$,
\begin{equation}
{d\log{\chi}\over d \log{t}}=-\gamma + B' t^{\theta'} .
\end{equation}
Thus the effective exponent depends on $t$ and, by scaling theory, on $L$,
\begin{equation}
\gamma_{\text{eff}} =\gamma + B'' t^{\theta'} = \gamma + C/L^{\theta} .
\end{equation}
Note that the same is true for the coefficient of a logarithmic singularity.

When studying lattice theories in the bulk, 
such as the $3$-dimensional classical Ising model, including or ignoring such conventional scaling corrections in the analysis of finite-size data
typically makes a difference of less than one percent in estimates of critical exponents such as $\gamma$.
For this reason, analyses of Monte Carlo simulation or series expansion results often ignore subleading corrections to finite-size scaling.
However, as first discussed by Cardy and Calabrese \cite{CardyCalabrese2010} in the context of $1+1$-dimensional conformal field theories (CFTs),
unusual corrections to scaling arise in the calculation of Renyi entropies.
This can be understood in the path-integral picture, where the entropy is proportional to the  
partition function of a $\alpha$-sheeted Riemann surface $\mathcal{R}_\alpha$
\cite{Holzhey:1994,Calabrese:2004} (Fig.~\ref{fig:conical_sing}), which at finite-temperature involves 
a branch cut located at $n \beta$ for $n=1,2,\cdots \alpha$ between each of the $\alpha$ replicas defined on region $B$.
A conical singularity exists on the boundary between regions $A$ and $B$ where this branch cut intersects the 
$\alpha \beta$-periodic imaginary time structure of the path integral in region $A$.
Local operators, which might even be \emph{relevant} in the bulk, are introduced at this conical singularity in
the multi-sheeted Riemann surface. Even though their contribution is only local, these operators yield subleading corrections to scaling with a power that
nevertheless diminishes with the Renyi index $\alpha$ and ultimately goes to zero as $\alpha$ goes to infinity, strongly altering the scaling behavior.

The unusual correction to scaling exponents in \mbox{$D=1+1$} are well known \cite{CardyCalabrese2010} 
and have been clearly observed  in numerical and analytical lattice calculations of spin chains \cite{UC1,UC2,UC3,UC4,UC5}.
For the case of the $D=1+1$ transverse-field Ising model, which we use as an example in the next section,
the most relevant operator that does not break the $Z_2$ symmetry is the energy operator with a scaling dimension $x_E=1$. 
For the Renyi correlation functions (defined in the next section) it leads to a finite-size correction to scaling 
like that of Eq.~(\ref{CHI_L}) but with an exponent,
\begin{equation}
\theta=x_E/\alpha, \label{theta-1d} 
\end{equation}
which goes to zero for large $\alpha$. 

While there is no theory for the unusual corrections to scaling in $D=2+1$, it is reasonable to expect that they are again
dominated by the presence of the energy operator at the conical singularity. The scaling dimension
of the energy operator $x_E$ is related to the correlation length exponent $\nu$ by the relation \cite{CardyBook},
\begin{equation}
x_E=d+1-{1/\nu}.   \label{higherDcorr}
\end{equation}
In particular, $x_E=1$ for the $D=2+1$ free boson theory, and approximately $1.41$ for the $2+1$ Ising theory \cite{Kos:2014}.
 In $D=2+1$ there is no particular reason why the correction should be given by the simple formula $x_E/\alpha$, as in $D=1+1$. 
 In Section \ref{3Dcorn}, we will see that while the exponent decreases with $\alpha$, the numerical results suggest that the dependence on $\alpha$ is slightly different.

\section{Renyi correlators in $D=1+1$} \label{RenyiCorrSection}

To illustrate a case where failing to account for the unusual corrections to scaling can lead to qualitative
changes in finite-size scaling, consider the transverse-field Ising model (TFIM) in spatial dimension $d=1$.  
The Hamiltonian is,
\begin{equation}\label{eq:ictf_hamiltonian}
{\cal H}= -h\ \sum_i \sigma_i^x -J\ \sum_{i} \sigma_i^z \sigma_{i+1}^z,
\end{equation}
where $\sigma^{x,y,z}$ are the Pauli operators. We focus on the critical point $h/J=1$, or its vicinity. 
The effect of unusual corrections to scaling has been well studied in this model for the Renyi entropies \cite{UC4} 
and related quantities \cite{Alba_2013,Coser_2014}.
Here, we instead examine the scaling of estimators weighted by the Renyi moments.
 One advantage of looking at these Renyi correlations is that one can observe more directly how the conical singularity affects
 correlations, not just the partition function (the entropy). As we will demonstrate, these correlation can easily be computed with conformal field theory techniques. 

To define the Renyi correlators, we divide the system into two halves $A$ and $B$ and
calculate the expectation value of operators (or pairs of operators) in the Renyi multi-sheeted geometry $\mathcal{R}_\alpha$.
The expectation value of an observable $\hat O$ in subsystem $A$ is defined as,
\begin{equation}\label{eq:renyicorrdef}
\langle \hat O\rangle_\alpha \ = \ {{\rm Tr} \ (\hat\rho_A)^\alpha \ \hat O \over {\rm Tr} \ (\hat\rho_A)^\alpha},
\end{equation}
where $\hat\rho_A$ is the reduced density matrix for subsystem $A$ and $\alpha$ is the Renyi index.
Note that setting $\alpha=1$ gives us normal bulk expectation values, which can also be
obtained without partitioning of the system, providing a non-trivial check on our computational procedures.

\subsection{Renyi correlators in conformal field theory}
We consider the simplest $D=1+1$ setup, which is a subsystem of length $\ell$ in an infinite system. As usual for such computations we use the replica trick: $\alpha$ is first assumed to be an integer, but the formulae are expected to hold for any $\alpha>0$.
 As explained in e.g. Refs.~\cite{Holzhey:1994,Calabrese:2004}, the entropy is -- up to some normalization -- nothing but the partition function of a $\alpha$-sheeted Riemann surface $\mathcal{R}_\alpha$. The Riemann surface may be mapped onto the complex plane by the conformal transformation,
 \begin{equation}\label{eq:conformal_map}
  z(w)=\left(\frac{w}{w-\ell}\right)^{1/\alpha}.
 \end{equation}
 Here, $w$ is a complex coordinate on the Riemann surface $\mathcal{R}_\alpha$, while $z$ is a complex coordinate in the complex plane $\mathbb{C}$. Subsystem $A$ corresponds to $w$ real in the interval $[0,\ell]$.

 To obtain a Renyi correlator in the ground state, what we need is to compute the correlation on $\mathcal{R}_\alpha$. Let us do that for some two point function of a primary operator. We use the two point function of a primary operator with scaling dimension $x$,
together with the transformation law of a primary operator \cite{BigYellowBook}. We obtain,
\begin{equation}\label{eq:intermediate}
 \Braket{\Phi(w)\Phi(w')}_{\mathcal{R}_\alpha}=\left|\frac{dz}{dw}\frac{dz'}{dw'}\frac{1}{(z-z')^2}\right|^x.
\end{equation}
Here $z$ (resp. $z'$) has to be understood as $z(w)$ (resp. $z(w')$). 
Using (\ref{eq:conformal_map}) this becomes,
\begin{equation}\label{eq:renyicorrcft}
 \Braket{\Phi(w)\Phi(w')}_{\mathcal{R}_\alpha}=\left|\frac{\ell}{\alpha w w'}\frac{\left(\frac{w w'}{(\ell-w)(\ell-w')}\right)^{\frac{1}{2}+\frac{1}{2\alpha}}}
 {\left(\frac{w}{\ell-w}\right)^{\frac{1}{\alpha}}-\left(\frac{w'}{\ell-w'}\right)^{\frac{1}{\alpha}}}
 \right|^{2x}\!\!.
\end{equation}
Eq.~(\ref{eq:renyicorrcft}) is the main result of this section. The Renyi correlations are translational invariant for $\alpha=1$, as they should be. This is also the case far from the boundaries $w=0,\ell$ for any $\alpha$. Similar to the entanglement entropy \cite{Calabrese:2004}, various generalizations to finite systems and finite temperatures are straightforward \footnote{For example the result for an interval of size $\ell$ in a periodic system of size $L$ follows from Eq.~(\ref{eq:intermediate}) with the conformal mapping $z(w)=\sqrt[\alpha]{\frac{\sin \pi w/L}{\sin \pi (\ell-w)/L}}$}.

In the following we are going to check our results in the Ising chain. The most natural example is the case when one of the points lies at the end of the interval $w_1=\delta$ ($\delta$ is a UV cutoff of the order of a lattice spacing; it ensures that the correlations stay finite), and the other at some distance $r$ from it:
\begin{eqnarray}
  \langle\Phi(\delta)\Phi(r)\rangle_{\mathcal{R}_\alpha}&\sim&\left|\frac{\ell}{\alpha \delta r} \frac{\left(\frac{\delta r}{\ell(\ell-r)}\right)^{1/2+1/2\alpha}}
  {\left(\frac{\delta}{\ell}\right)^{1/\alpha}-\left(\frac{r}{\ell-r}\right)^{1/\alpha}}\right|^{2x}\\\label{eq:simpletest}
  &\sim& \frac{\left(1-\frac{r}{\ell}\right)^{x/\alpha-x}}{r^{x/\alpha+x}}\left(A_\alpha +O(r^{-\frac{1}{\alpha}})\right),
\end{eqnarray}
where $A_\alpha$ is some $\alpha$-dependent UV cutoff. The correction terms $O(r^{-1/\alpha})$ lead to the correction to scaling exponent
$1/\alpha$ in Eq.~(\ref{theta-1d}).
 We are going to test formula (\ref{eq:simpletest}) in the following subsections. Note also, we recover that the scaling dimension of an operator near the cone is modified to $x\to x/\alpha$ (Eq.~(\ref{theta-1d})), as pointed out in Ref.~\cite{CardyCalabrese2010}.
 
\subsection{A numerical check using free fermions}\label{sec:freefermions}
It is well known that the Hamiltonian (\ref{eq:ictf_hamiltonian}) can be mapped onto a system of free fermions through a Jordan-Wigner transformation,
\begin{align}\label{eq:JW1}
 \sigma_j^x&=2c_j^\dag c_j-1,\\\label{eq:JW2}
 2\sigma_j^+&=\sigma_j^z+i\sigma_j^y=2c_j^\dag \exp\left(i\pi \sum_{l<j}c_l^\dag c_l\right),
\end{align}
and then diagonalized by a Bogoliubov transformation. The correlations in the ground state may then be obtained analytically \cite{LSM}.

It is also well known how to compute the entanglement entropy for such quadratic fermion systems \cite{ChungPeschel,Peschel}. The method relies on the following observation. Since Wick's theorem holds for any fermionic correlation in subsystem $A$, the reduced density matrix (RDM) itself is the exponential of a quadratic fermion Hamiltonian $H_A$, $\rho_A=e^{-H_A}/{\rm Tr}\, e^{-H_A}$. Finding the precise form of $H_A$ can then be done by demanding that it reproduce all correlation functions in subsystem $A$, $\braket{\hat{O}}={\rm Tr\,}(\rho_A \hat{O})$.  
To perform the identification one of the easiest way is to use the correlation matrix,
\begin{equation}
 K=\left(\begin{array}{cc}
          C&D\\D^*&1-C^t
         \end{array}
\right),
\end{equation}
where $M^t$ (resp. $M^*$) denotes the transpose (resp. conjugate transpose) of any matrix $M$. The $\ell\times \ell$ matrices $C$ and $D$ encode the two types of correlators:
\begin{eqnarray}
 C&=&(\braket{c_i^\dag c_j})_{1\leq i,j\leq \ell},\\
 D&=&(\braket{c_i^\dag c^\dag_j})_{1\leq i,j\leq \ell}.
\end{eqnarray}
The eigenvalues of $K$ are simply related to the single particle eigenenergies of the entanglement Hamiltonian \cite{Peschel}, and the knowledge of the former allows to determine the latter. Said differently, given a subsystem correlation matrix it is easy to determine the corresponding entanglement Hamiltonian, and then to compute the entropy.

Here we want to compute the Renyi correlations, as defined in Eq.~(\ref{eq:renyicorrdef}). Since the RDM is the exponential of a quadratic form, so is any power of it. The new entanglement Hamiltonian corresponding to the new RDM $\rho_A^\alpha/{\rm Tr}\rho_A^\alpha$ is then simply $\alpha H_A$. One can then find the modified Renyi correlation matrix $K_{\alpha}$ that would have $\alpha H_A$ as the entanglement Hamiltonian. The details are explained in Appendix~\ref{app:freeboson}. In compact form the result may be written,
\begin{equation}\label{eq:renyifreefermions}
 K_\alpha=\left[1+\left(K^{-1}-1\right)^\alpha\right]^{-1}.
\end{equation}
Note that in absence of pairing terms $c^\dag c^\dag$, the same formula holds with $K$ and $K_\alpha$ replaced by $C$ and $C_\alpha$. Using (\ref{eq:renyifreefermions}), one can access all Renyi correlations between fermions, $\braket{c_i^\dag c_j}_\alpha$, $\braket{c_i^\dag c_j^\dag}_\alpha$, the spin correlations in the $z$ basis being recovered by using the Jordan Wigner transformation (\ref{eq:JW1}, \ref{eq:JW2}), and appropriate treatment of the resulting Jordan-Wigner string (see e.g. Ref.~\onlinecite{LSM}). For the spin-spin correlation  at distance $r$ we obtain,
\begin{equation}\label{eq:szszdet}
 \braket{\sigma_i^z \sigma_{i+r}^z}_\alpha=\det_{1\leq m,n\leq r}\left(\braket{a_{i+m-1}b_{i+n}}_\alpha\right),
\end{equation}
with $a_j=c_j^\dag-c_j$ and $b_j=c_j^\dag+c_j$.

\begin{figure}[htbp]
  \includegraphics[width=0.48\textwidth]{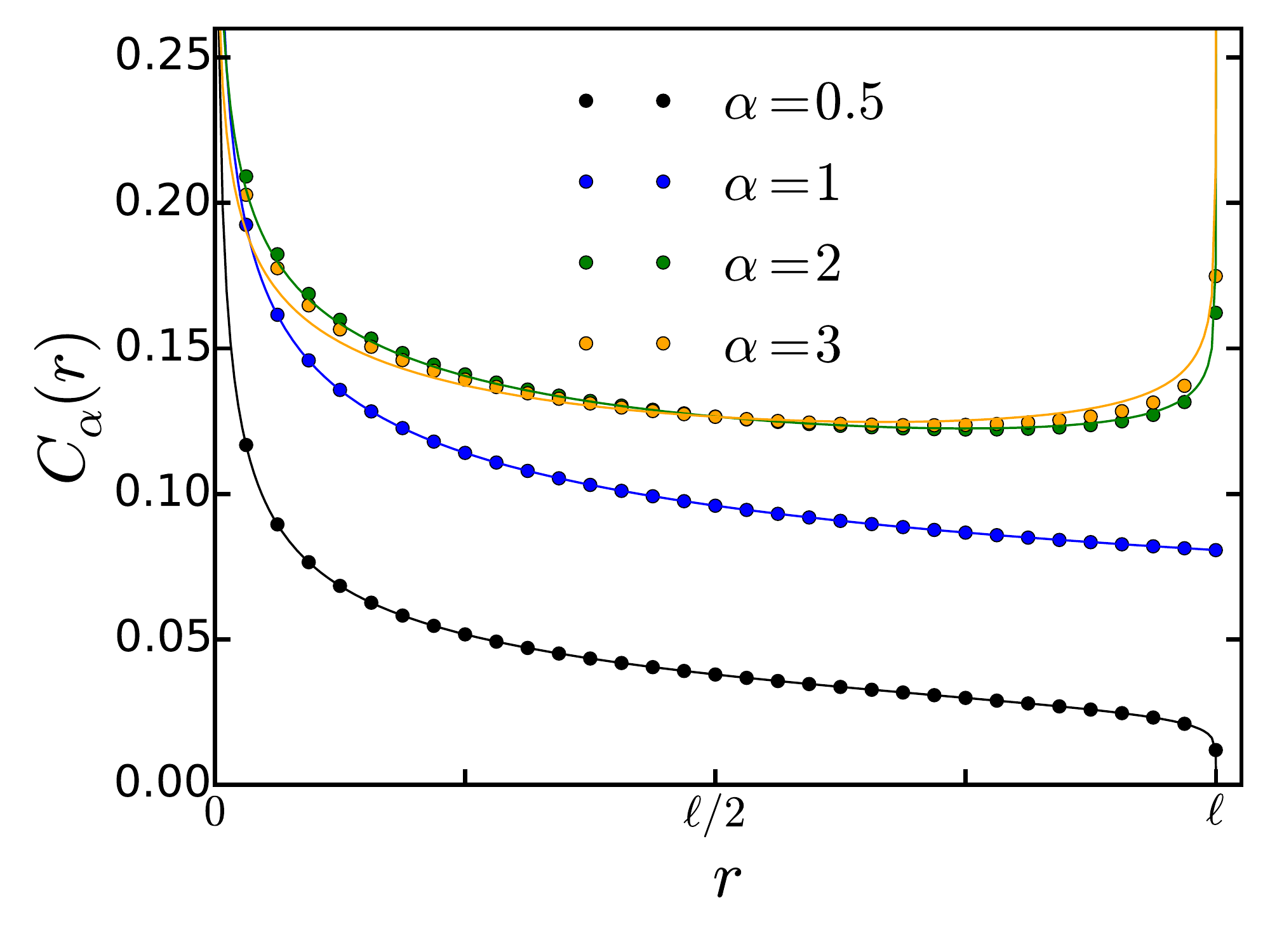}
  \caption{Renyi spin-spin correlation $C_\alpha(r)=\braket{\sigma_{0}^z \sigma_{r}^z}_\alpha$ in the Ising chain for several values of $\alpha=0.5,1,2,3$. The full system is infinite, and the subsystem size is $\ell=4096$. The dots are the numerical data, while the lines are the CFT result. The UV cutoff $A_\alpha$ in (\ref{eq:simpletest}) is adjusted such that the analytical result matches the numerical one at $r=\ell/2$.}
  \label{fig:Isingfermions}
\end{figure}

We now use (\ref{eq:szszdet}) to numerically compute the correlation,
\begin{equation} \label{Calpha}
C_\alpha(r)=\braket{\sigma_0^z \sigma_r^z}_\alpha, 
\end{equation}
in the ground state of the chain (\ref{eq:ictf_hamiltonian}) at the critical point $h/J$=1. 
We consider several values of $\alpha$ and several distances $0\leq r\leq \ell$ in a subsystem of length $\ell=4096$. As can be seen in Fig.~\ref{fig:Isingfermions}, the agreement with the prediction (\ref{eq:simpletest}) with Ising/Onsager exponent $x=1/8$ is very good. Note however that finite-size effects increase significantly with $\alpha$; indeed the formula is expected to be valid only in the regime $r^{1/\alpha}\gg 1$, which for moderately large values of $\alpha$ already requires immense system sizes. 
Another interesting feature is that the correlation near $r=\ell$ blows up again at $r=\ell$ for $\alpha>1$. One possible interpretation would be that since the scaling dimension $x$ near the cone is modified to $x/\alpha$, the spin configurations near the (fictitious) endpoints at $r=0,\ell$ are more constrained, so more likely to match. Since translational invariance is broken, the effect appears similar to that of a boundary with partial reflection. 

Of course, the fact that we are able to test our results with good precision relies heavily on the free fermion structure. For more general system the computation of Renyi correlations is more difficult, and the accessible system sizes are much smaller. To illustrate this we look at similar correlations with a different method below. 

\subsection{Correlation Function and Structure Factors}
\begin{figure*}[ht]
\includegraphics[width=\columnwidth]{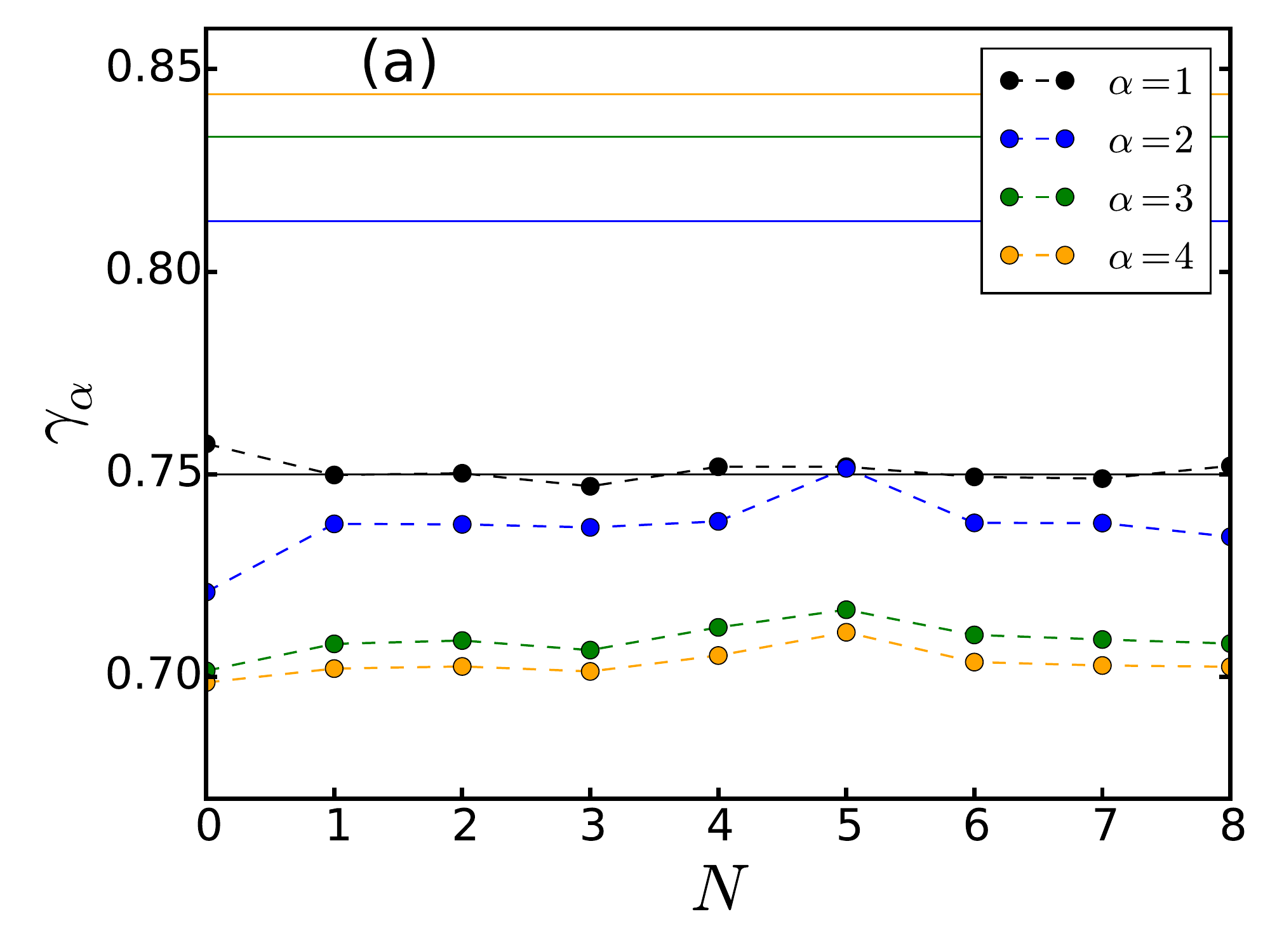}
\includegraphics[width=\columnwidth]{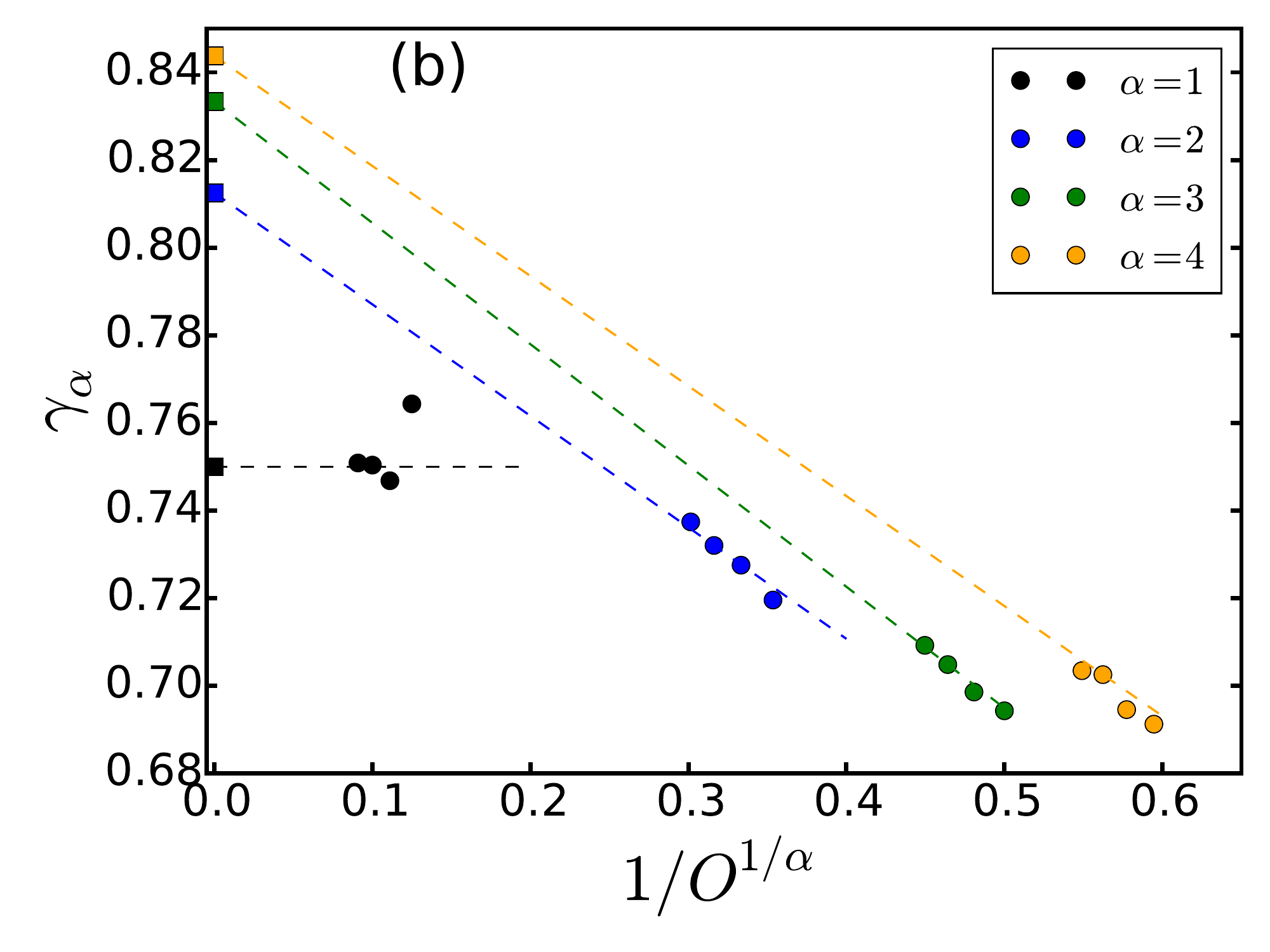}
\caption{(a) Structure factor exponents estimated with the full length series.  The $N$-axis labels different approximates.  The solid circles are the estimated values (with dashed lines as a guide to the eye).  The solid lines are the theoretically predicted values.
(b) Estimated exponents for different orders $O$ of the series expansion.  The squares are the asymptotic theoretical values.  Dashed lines are guides to the eye, showing that a rough extrapolation of data is consistent with theory.}
\label{fig:SF}
\end{figure*}

Let us calculate correlation functions $C_\alpha(r)$ directly in the high-field (disordered) phase 
of Eq.~(\ref{eq:ictf_hamiltonian}) by a series expansion in $J/h$ \cite{oitmaa}. 
We define
$C_\alpha(r)$ via Eq.~(\ref{Calpha}),
where $0$ is a boundary site of $A$ directly at the interface with $B$ and $r$ refers to any
other site in subsystem $A$. 
For correlation function at $r$ in Eq.~(\ref{Calpha}), 
the series coefficients will be zero below order $r$. One needs at least $r$ powers of the perturbation
to get a non-zero correlation between site $0$ and site $r$.
This means as one goes further away in distance, the series become effectively shorter and shorter. For this reason,
analyzing real-space correlations is not very useful. Series extrapolation methods, which are necessary for
studying the critical region, simply cannot be applied. Instead, we need to focus on structure factors
or correlation sums defined as,
\begin{equation}
{\mathcal S}_\alpha = \sum_{r} C_\alpha(r) \label{StructF}.
\end{equation}
It is useful but not necessary to multiply all $r>0$ terms in the sum in Eq.~(\ref{StructF}) by a factor of 2, as that would
be the equal-time structure factor of the 1D system in the absence of a partition, where one has two neighbors
at each such distance. This is what we have done in the analysis discussed below.

We describe the series analysis using the method of differential approximants \cite{fisher,baker} in Appendix \ref{AppA}.  There, we 
define $O=M+N+J+1$ as the order of the series used in the
estimation of the critical exponent ($M$, $N$, and $J$ are individually the orders of three polynomials defining the extrapolation scheme). 
In other words, $O$ is the maximum power used in determining the polynomial coefficients.
The order of the series expansion is related to a length scale \cite{Affleck:1990}, thus effective exponents can vary with the order as,
\begin{equation}
\gamma_{\text{eff}} =\gamma + C/O^{\theta}.
\end{equation}
Again, theory predicts that the scaling dimension of the bulk spin is given by, 
$x_{bulk}=1/8$.
The scaling dimension of the boundary spin with Renyi index $\alpha$ is given by,
\begin{equation}
x_{\alpha}=\frac{x_{bulk}}{\alpha}={1\over 8\alpha}.
\end{equation}
Thus the correlation function should decay for large $r$ as,
\begin{equation}\label{eq:bulkplusbound}
C_\alpha(r)= A/r^{x_{bulk}+x_{\alpha}}.
\end{equation}
Summing over $r$, the structure factor ${\mathcal S}_\alpha$ should diverge on approach
to the critical point with an exponent,
\begin{equation}
\gamma_\alpha=(1-x_{bulk}-x_\alpha)\nu.
\end{equation}
Since, $\nu=1$ for the model, the exponent simplifies to,
\begin{equation}
\gamma_\alpha=(1-x_{bulk}-x_\alpha)= (1-{1\over 8}-{1\over 8\alpha}).
\end{equation}

Figure \ref{fig:SF}(a) shows the estimated structure factor exponent obtained from order $(J/h)^{11}$ series expansions with various choices for 
the order $N$ of one of the polynomials used in the fit.  
Note that the exponent $\gamma$ is converged very well to the theoretical value for $\alpha=1$ (which gives the bulk exponent), 
but the convergence is poor for larger~$\alpha$. 
In fact, while theory predicts that the exponent $\gamma$ should increase with $\alpha$, the series extrapolation estimate appears to systematically decrease.

Figure \ref{fig:SF}(b) shows plots of the estimated exponent values for different orders of the series. The
$x$ axis has been scaled differently for different $\alpha$ values. The rescaling pushes the data
for larger $\alpha$ values more and more to the right. Thus for larger $\alpha$ one needs
to extrapolate farther to reach the asymptotic value. The solid squares are the asymptotic theoretical
values and the dashed lines are a guide to the eye showing the results are clearly consistent with
theory, only when the unusual corrections to scaling are taken into account.

\subsection{Magnetization in the ordered phase}
\begin{figure*}[htbp]
\includegraphics[width=\columnwidth]{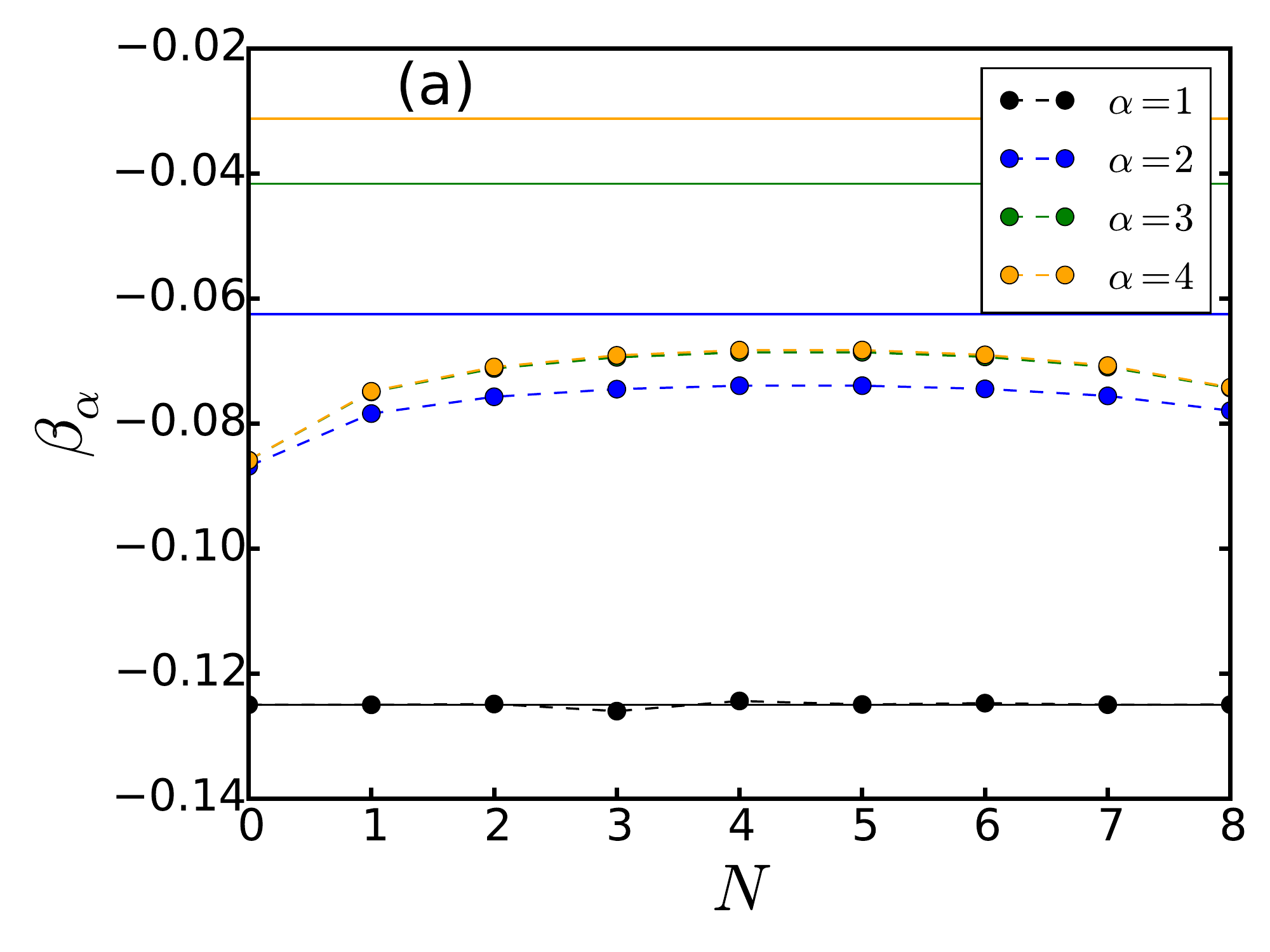}
\includegraphics[width=\columnwidth]{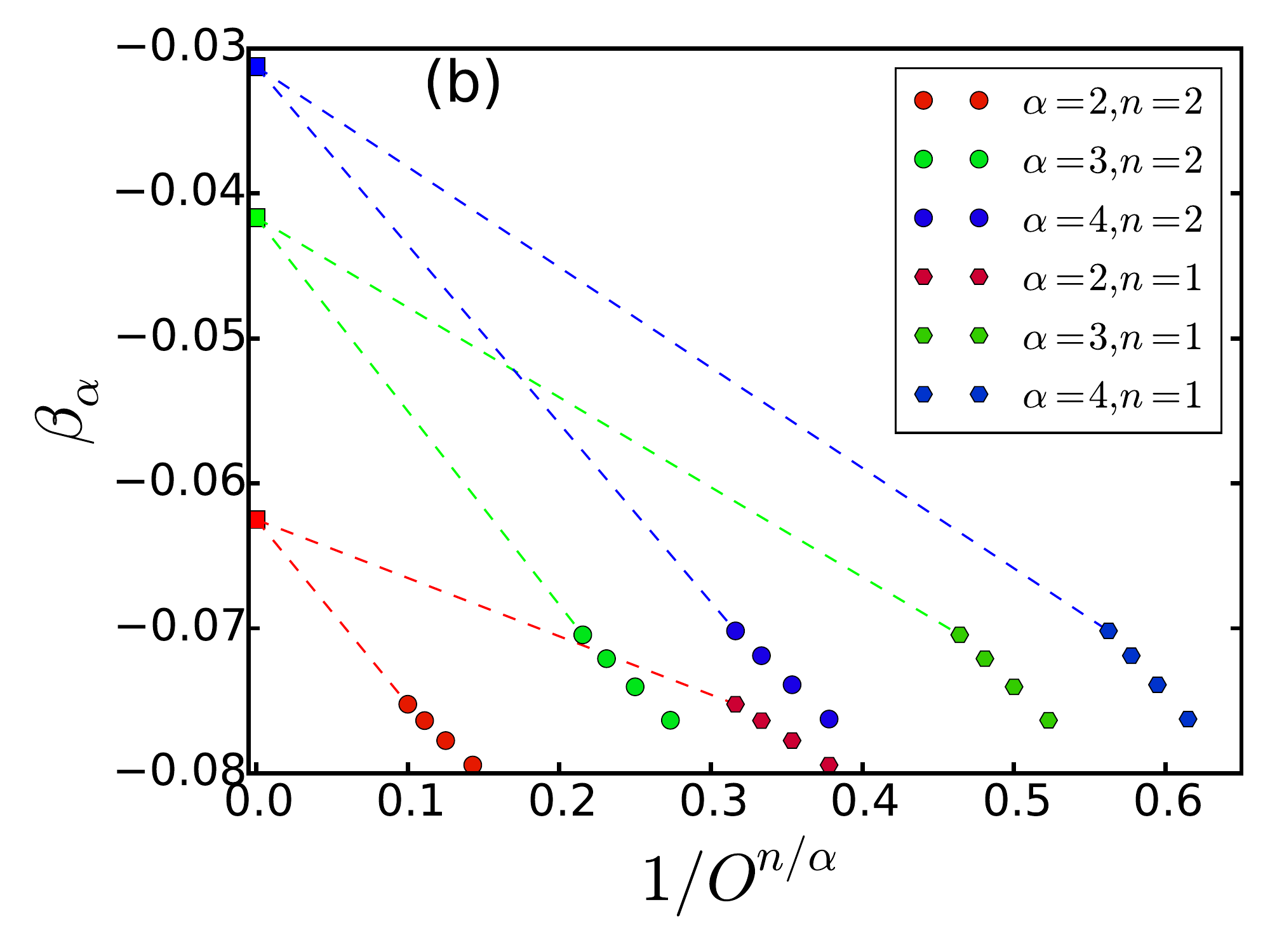}
\caption{(a) Magnetization exponents estimated with the full length series.  The $N$-axis labels different approximates.  The solid circles are the estimated values (with dashed lines as a guide to the eye).  The solid lines are the theoretically predicted values.
(b) Estimated exponents for different orders $O$ of the series expansion.  The squares are the asymptotic theoretical values.  Dashed lines are guides to the eye, showing that a rough extrapolation of data is consistent with theory.}
\label{fig:M}
\end{figure*}
We now calculate series expansions for the Renyi magnetization at the boundary site of $A$ by a series expansion in $h/J$
(a low-field expansion in the ordered phase). This boundary magnetization is defined as,
\begin{equation}
M^0_\alpha=\langle \sigma^z_0\rangle_\alpha.
\end{equation}
Once again, setting $\alpha=1$ just gives us the usual bulk magnetization that is well known to vanish at the
critical point with exponent $\beta=1/8$.

When extrapolating, we do not allow any $P_3^J(x)$ terms since the magnetization vanishes identically at the critical point with no background value (see Appendix \ref{AppA}). 
Without such terms, these approximants are also known as d-log Pade approximants. 
The order $O$ now equals $M+N$.

The boundary magnetization vanishes with the scaling dimension of the boundary operator (with $\nu=1$)
\begin{equation}
\beta_\alpha={1\over 8 \alpha}.
\end{equation}
Figure \ref{fig:M}(a) and \ref{fig:M}(b) show corresponding results calculated up to order $(h/J)^{20}$ for the
magnetization. Note, however, that the magnetization series only has even powers, so it is effectively a
ten-term series in $(h/J)^2$.

Notice that for the magnetization, the variation of the estimated exponents
with order appears to adhere more closely to a $O^{2/\alpha}$ dependence.
Such terms are always present in addition to the $O^{1/\alpha}$ terms, but should usually be weaker.
However, we must use even more caution in extrapolating the magnetization exponents
to asymptotic values, especially for $\alpha>2$, as our data is still very far from the final values.
\section{Renyi entropies in $D=2+1$ corners} \label{3Dcorn}

Having seen the importance of unusual corrections to scaling for Renyi quantities in space-time dimension $1+1$, 
we turn to a case of intense recent interest: universal Renyi entropy corrections for $D=2+1$ critical systems.
In two spatial dimensions, universal quantities appearing in the Renyi entropies have a rich dependence on subregion geometry.
For example, various universal terms (sub-leading to the area law) can arise from geometries such as smooth circular bipartitions \cite{Casini:2009,Casini:2012}; sharp corners \cite{Casini:2007,CornerCQCP1,CornerCQCP2,CornerCQCP3}; bipartitioned infinite cylinders \cite{Metlitski:2009}
or cylindrically-bifurcated tori \cite{Ju_2012,Stephan_2013,Inglis_2013,Fradkin_2015}.

Significant computational effort has been dedicated recently toward calculating the subleading contributions to the 
entanglement entropy arising from $\pi/2$ corners in quantum models on the square lattice. Here we study the free scalar field theory, as well as the two-dimensional Ising chain in transverse field. 
A bipartition with a single corner (e.g.~a quadrant of the infinite plane) in the entangling boundary contributes a subtractive, divergent logarithmic 
correction to the Renyi entropy,
\begin{equation}
S_{\alpha} = C_{\alpha} \frac{\ell}{\delta} - a_{\alpha}(\pi/2) \log \left({ \frac{\ell}{\delta} }\right) + \cdots, \label{EEscale}
\end{equation}
where $\ell$ is the linear length scale of the bipartition $A$,
and $\delta$ is the UV cutoff corresponding to the lattice length-scale.
The constant $a_{\alpha}$ is not polluted by UV details, and constitutes a universal number that can be explored
easily for $\pi/2$ using lattice numerics.

Previous numerical studies reveal that the value of $a_{\alpha}(\pi/2)$ obtained for the $O(N)$ model is 
somewhat close to $N$ times the value obtained for a free scalar field theory by Casini and Huerta \cite{Casini:2007}.   
In this section, we use the numerical linked-cluster expansion (NLCE) to examine the finite-size scaling 
behavior of $a_{\alpha}(\pi/2)$ for free bosons, and perform a parallel comparison of data obtained for the
transverse field Ising model (TFIM) in two spatial dimensions, at its $N=1$ fixed-point. 
For the latter we use the density matrix renormalization group (DMRG) 
as our cluster solver \cite{Schollwoeck:2005,ITensor}.

In our NLCE procedure, described in detail in Appendix \ref{AppNLCE}, the leading-order area-law piece
of Eq.~(\ref{EEscale}) cancels, leaving only the logarithmic divergence,
\begin{equation}
{\mathcal V}_{\alpha} = - a_{\alpha} \log \ell + b_{\alpha}. \label{eq:Corner}
\end{equation} 
In the next two sections, 
we interpret the order ${O}$ of the NLCE (see Appendix \ref{AppNLCE}) as a length scale $\ell$,  a procedure used successfully several times in the past \cite{Kallin:2013,Kallin:2014,Miles:2014}.
This length scale is used to fit the two parameters of Eq.~(\ref{eq:Corner}) to extract the universal constant $a_{\alpha}(\pi/2)$.

\begin{figure*}[t]
\includegraphics[width=\columnwidth]{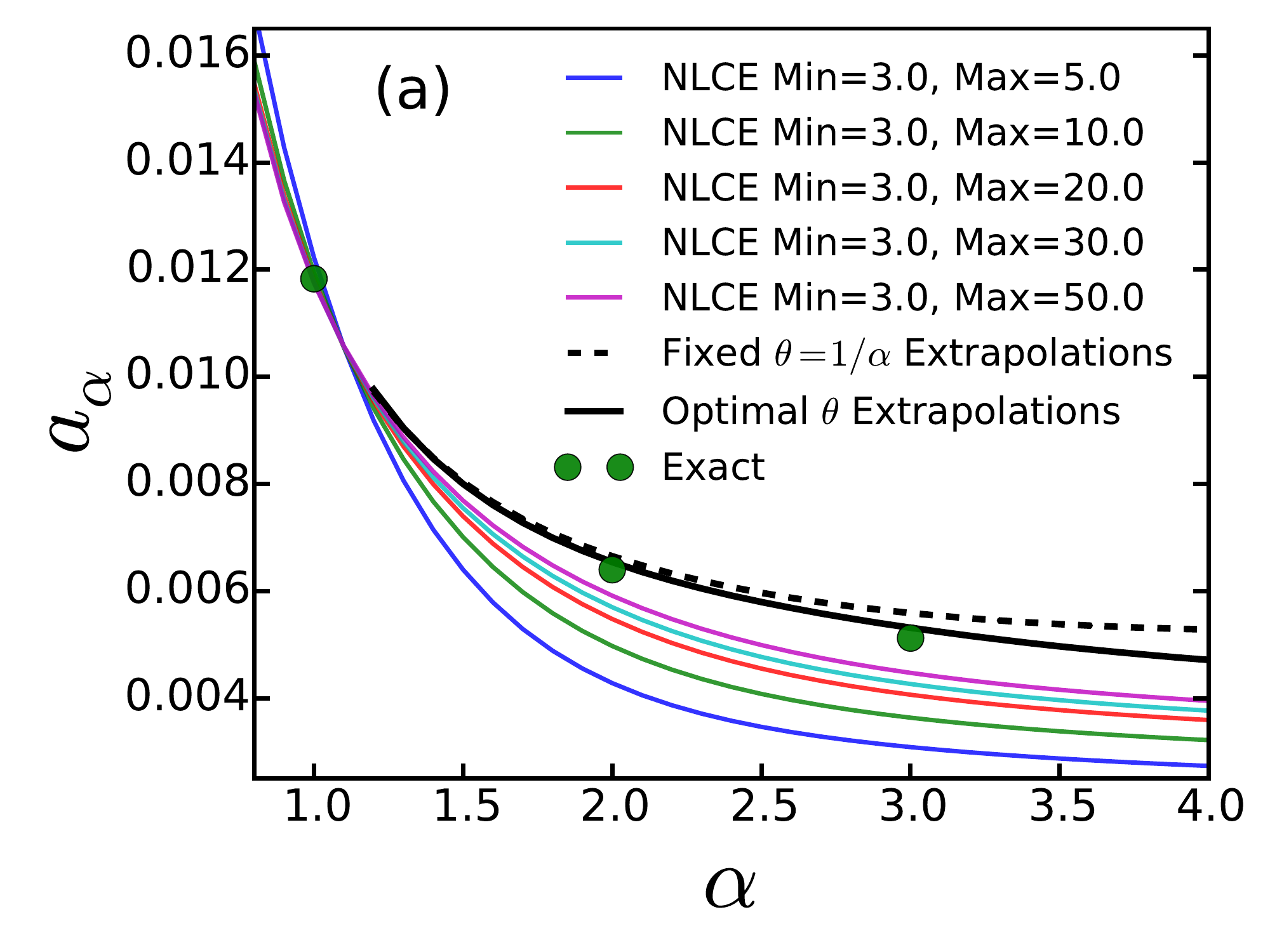}
\includegraphics[width=\columnwidth]{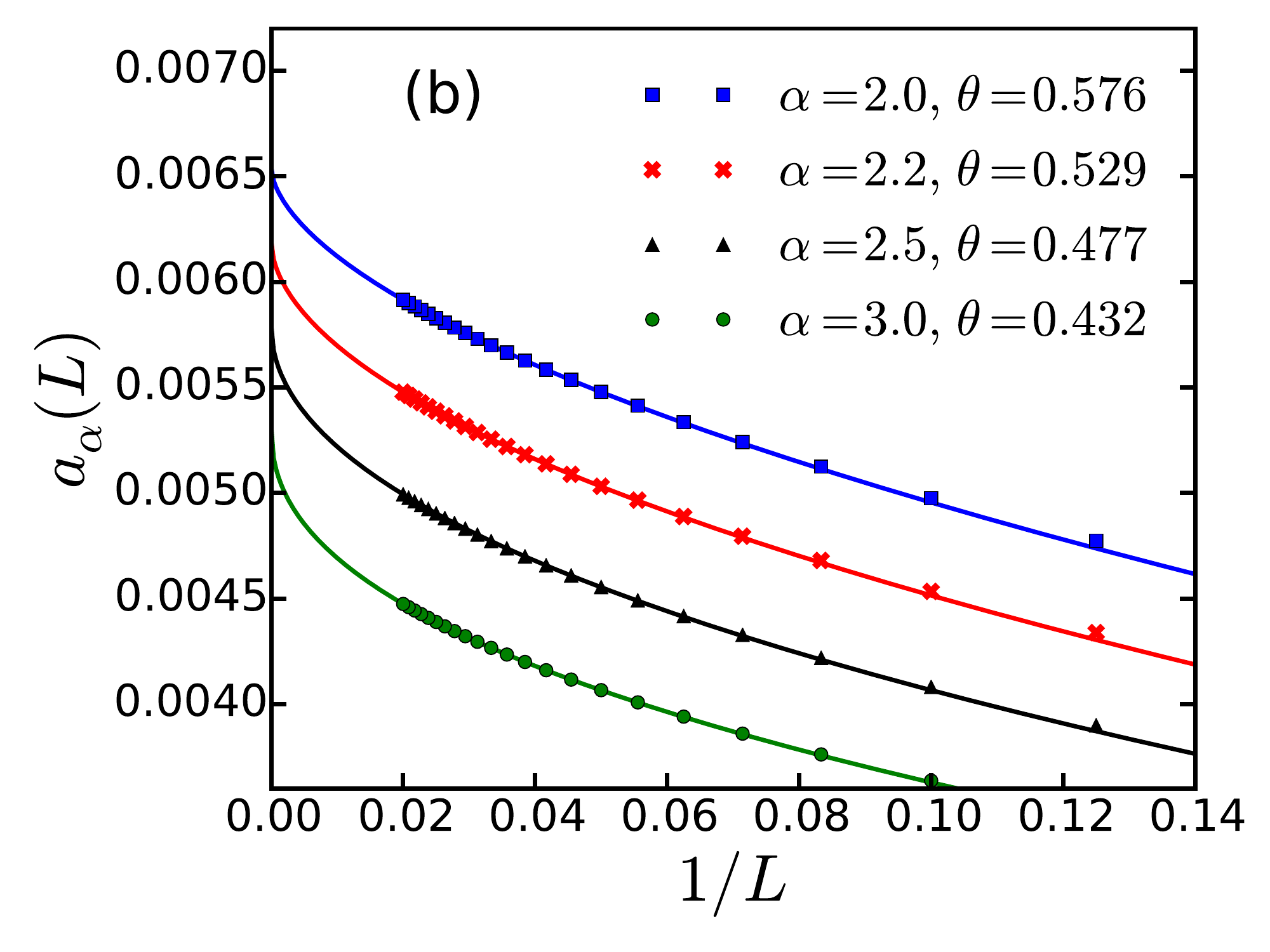}
\caption{(a) Numerical linked cluster results for the universal corner coefficient  of the free scalar field theory, as a function of Renyi index $\alpha$. Each curve was obtained by 
fitting NLCE data to the form Eq.~(\ref{eq:Corner}), using only data between the minimum order and maximum order indicated. 
(b) Two Renyi coefficients, obtained by fitting NLCE results from order 3 to order $L$, as a function of maximum order $L$.  
Solid curves are fits to $A + B/L^\theta$, with coefficients in the key obtained by fitting the largest 10 $L$ points.
}
\label{fig:bosons1}
\end{figure*}

\subsection{Free scalar field theory}\label{sec:fsft}

We begin by employing the NLCE method to  study the free scalar field theory, described in the Appendices.
For the free boson, the NLCE procedure can be used with a correlation-matrix technique as the cluster solver, 
as detailed in Appendix \ref{app:freeboson}.  The correlation-matrix technique permits the calculation of 
Renyi entropies on very large finite-size clusters, allowing the NLCE sum to be carried to extremely high order.

In Fig.~\ref{fig:bosons1}(a), we extract the universal coefficient $a_{\alpha}(\pi/2)$ from NLCE data over a range of orders, from ${O}_\text{min}$ to ${O}_\text{max}$,
by fitting Eq.~(\ref{eq:Corner}) as a function of NLCE order with two free parameters ($a_{\alpha}$ and  $b_{\alpha}$).
For reference,  Fig.~\ref{fig:bosons1}(a) includes the exact values for the thermodynamic limit  obtained by Casini and Huerta \cite{Casini:2007}.
The von~Neumann corner coefficient $a_{1}(\pi/2)$ converges quite rapidly to the exact value, while the Renyi coefficients $a_{2}(\pi/2)$ and $a_{3}(\pi/2)$ 
converge much more slowly, even for very large linear cluster sizes $O_{max}=50$.

As discussed in Section \ref{UnusualSection}, one may expect that for $\alpha > 1$ a second extrapolation could become necessary because of the 
unusual corrections to scaling arising from the conical singularities.  
We perform this second extrapolation in Fig.~\ref{fig:bosons1}(b).
The individual points are obtained by first fitting the NLCE data as a function of order 
to Eq.~(\ref{eq:Corner}) from $O_\text{min}$ up to a fixed maximum order ${O}_\text{max}$.  This is the same procedure discussed above for producing the curves labeled 
NLCE in Fig.~\ref{fig:bosons1}(a), but keeping $\alpha$ fixed and combining results for various ${O}_\text{max}$ into a single curve. 
Next, we identify each $O_\text{max} \equiv L$, a length scale which depends on the NLCE scheme (see Appendix \ref{AppNLCE}).
We then fit the finite-size results to the function $A + B/L^\theta$ with $A, B$ and $\theta$ as fitting parameters.  
From Eq.~(\ref{higherDcorr}), taking \mbox{$x_E=1$} for $D=2+1$, 
one might guess the correction should go as $L^{-x_E/\alpha}$, i.e.~$\theta=1/\alpha$, in analogy to the $D=1+1$ result.
However, as we see in Fig.~\ref{fig:bosons1}(b), the optimized values of $\theta$ which produce the best fit are somewhat different
from $1/\alpha$ and yield extrapolations closer to the exact results \cite{Casini:2007},
likely indicating different scaling exponents associated with the higher-dimensional system.

From these extrapolations, we estimate the following values for $a_{\alpha}(\pi/2)$ in the thermodynamic limit:
for $\alpha = 2$ we obtain \mbox{$a_2(\pi/2) = 0.0065$} which is 1\% off of the exact value $0.0064$ of Ref.~\cite{Casini:2007}; for $\alpha=3$ we obtain \mbox{$a_3(\pi/2) = 0.0053$} which is 3\% from the exact value of $0.0051$.
Clearly, our NLCE estimates will approach the exact results 
with increasing accuracy as $O_\text{max}$ increases to infinity.

\subsection{Transverse-field Ising model}

\begin{figure*}[ht]
\includegraphics[width=\columnwidth]{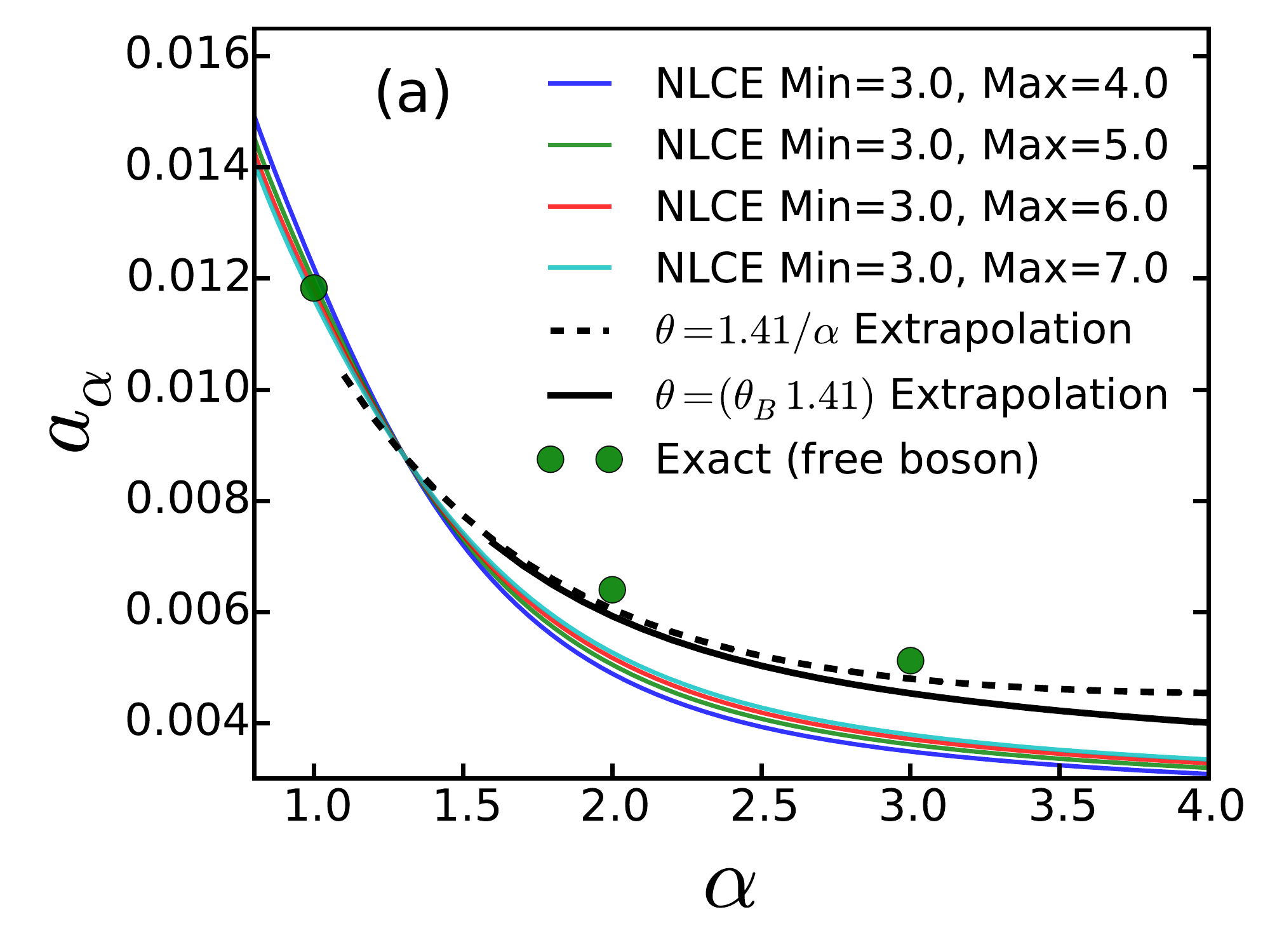}
\includegraphics[width=\columnwidth]{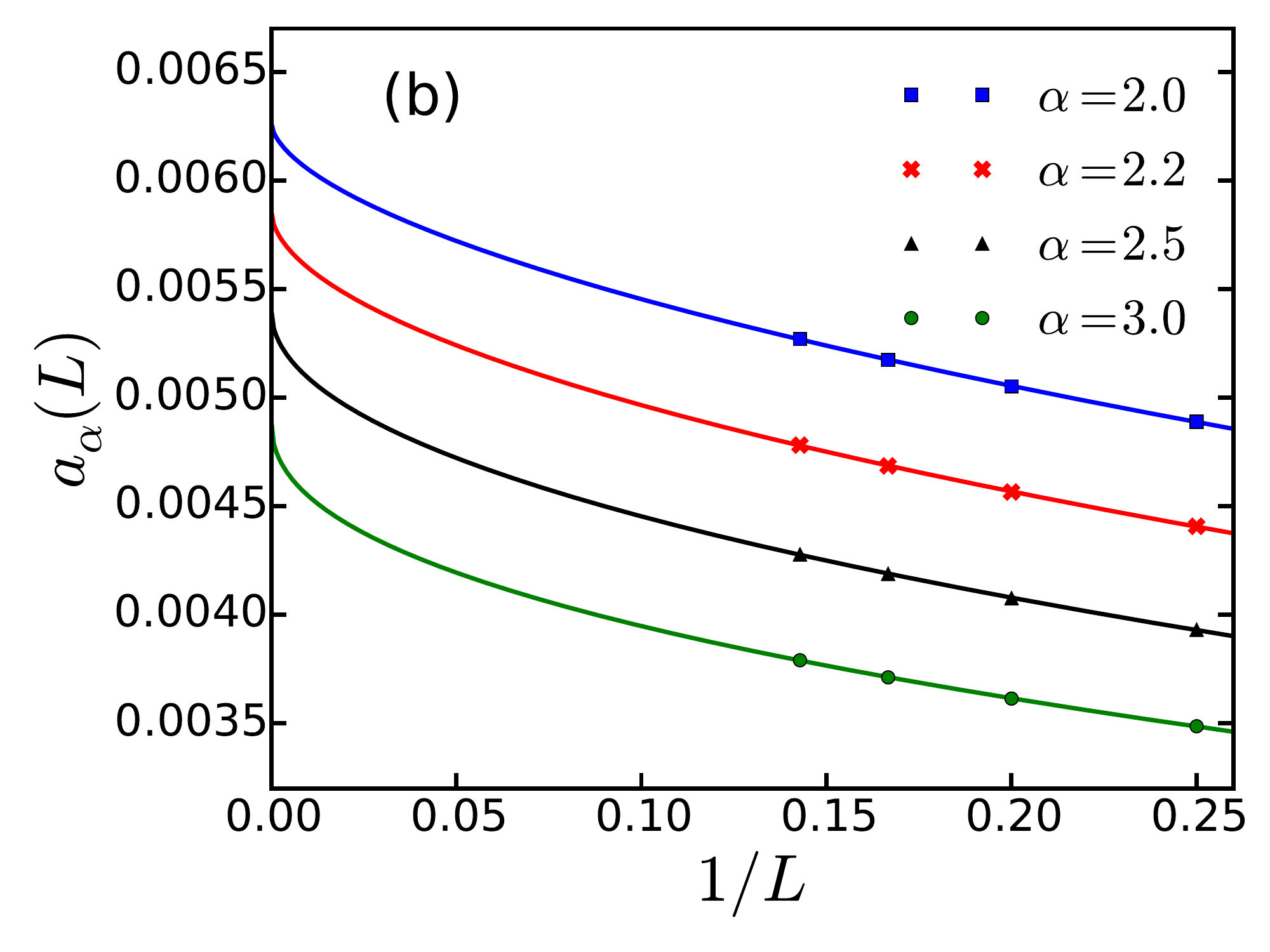}
\caption{(a) Numerical linked cluster results for the universal corner coefficient  at the quantum critical point of the TFIM, as a function of Renyi index $\alpha$. The lower curves were obtained by 
fitting NLCE data to the form  Eq.~(\ref{eq:Corner}), using only data between the minimum order and maximum order indicated. The two curves 
labeled ``Extrapolation'' were obtained using the two-step fitting procedure described in the text, in one case fixing the power-law exponents to be $\theta=1.41/\alpha$ and in the 
other case $\theta=1.41 \cdot \theta_B$ where $\theta_B$ is the optimal exponent for fitting the finite-size free boson results.
(b) Finite-size Renyi corner coefficients $a_\alpha$ estimated fitting NLCE data up to some maximum order $L$. Solid curves are fits to $A + B/L^{\theta}$ with $\theta=1.41 \cdot \theta_B$. The intercepts at $1/L \rightarrow 0$
define the solid curve labeled ``Extrapolation'' in subfigure (a).}
\label{fig:tfim1}
\end{figure*}

We turn now to NLCE calculations of the transverse-field Ising model (TFIM) in spatial dimension $d=2$.  The Hamiltonian on a square lattice is
\begin{eqnarray}
{\cal H}&=& -h\ \sum_i \sigma_i^x -J\ \sum_{\langle ij \rangle} \sigma_i^z \sigma_{j}^z,
\end{eqnarray}
and we work at the quantum critical point \mbox{$h/J \approx 3.044$}.
The corner contribution to the Renyi entropy has been studied by NLCE several times in the past  \cite{Kallin:2013,Kallin:2014}.  The main technical difference 
from the free boson calculation of the preceding section is that each individual cluster must be solved using a method for strongly correlated
Hamiltonians.  This severely restricts the value of $O_{max}$, even if one uses the powerful DMRG method as a cluster solver.

In Fig.~\ref{fig:tfim1},  the universal coefficient $a_{\alpha}(\pi/2)$ is extracted from NLCE data in a range of orders, from $O_{min}$ to $O_{max}$,
using the fitting form Eq.~(\ref{eq:Corner}) with two free parameters ($a_{\alpha}$ and  $b_{\alpha}$).
Up to the maximum order we reach, the results obtained this way for the TFIM are significantly different from the exact results for the free boson, particularly for 
$\alpha \ge 2$.
However, if one performs a second finite-size extrapolation for the Renyi coefficients, using the form $A+B/L^\theta$ with $\theta$ values
taken to be $\theta_\text{TFIM} = 1.41 \times \theta_B$, where $\theta_B$ are the optimal exponents 
for the free boson and $x_E\approx 1.41$ is the TFIM energy operator scaling dimension Eq.~(\ref{higherDcorr}), then our estimates for $a_{\alpha}(\pi/2)$ in the limit $L\rightarrow \infty$ 
are 0.0059 for $\alpha=2$; 0.0045 for $\alpha=3$; and 0.0040 for $\alpha=4$. As shown in Fig.~\ref{fig:tfim1}(a), if we instead
assume the exponents $\theta_\text{TFIM} = 1.41/\alpha$, our estimates are 0.0061 for $\alpha=2$; 0.0048 for $\alpha=3$; and 0.0045 for $\alpha=4$, 
higher than the extrapolation based on the boson exponents, especially for larger $\alpha$.

The data indicate that a large part of the observed difference between free bosons and the TFIM in past studies 
may be the unusual corrections to scaling, which are manifest as 
large finite-size effects in the NLCE procedure for $\alpha \neq 1$.
When such unusual corrections, arising from the presence of the conical singularity in the Renyi entropies, are taken into account,
it becomes much more difficult to distinguish the value of $a_{\alpha}(\pi/2)$ obtained for the TFIM from that for free bosons.  Since it will be very 
difficult to obtain larger values of $O_{max}$ using the DMRG method, it is hard to judge how much of the remaining discrepancy can be
attributed to the very limited cluster sizes of the TFIM NLCE calculation.

\subsection{Discussion}

In this paper, we have performed a systematic study of the finite-size scaling of Renyi entropies and Renyi correlators at a quantum critical point. In particular, special attention was paid to the first subleading correction, which can play a crucial role in a proper extraction of universal terms.

In space-time dimension 1+1, a known scaling ansatz has been derived by Calabrese and Cardy and applied previously
to numerical studies of Renyi entropies.  We have applied this ansatz to
Renyi correlators, calculated with conformal field theory, a mapping to free fermions, and series expansions for the transverse-field Ising model (TFIM)
at its quantum critical point.
For data obtained with finite-size (or finite-series length) calculations, we find that 
a second extrapolation in the size (or length of the series) using this ansatz is needed to show consistency with the theoretically known values.
Failing to take the correction to scaling into account can lead to qualitatively incorrect answers.
Thus, it appears that the conical singularities are far more important for Renyi correlators in $D=1+1$ than
irrelevant operators are, e.g.~for computing bulk properties of the $3$-dimensional classical Ising model.

We also studied $2+1$ space-time dimensions, where the influence on scaling of the conical singularity is generally not known exactly.
We focus on the universal term $a_\alpha$ which arises in Renyi entropies due to a $\pi/2$ corner, using
a recently developed numerical linked cluster expansion (NLCE). 
In a scalar field-theory corresponding to free bosons, where the exact result is known, very large systems can
be studied.  
As in 1+1, we see that strong corrections to scaling are present for $\alpha>1$ Renyi entropies, 
which retain a significant discrepancy from exact results even for relatively large cluster sizes.
However, if we do an additional extrapolation in the maximum linear dimension as $L^{-\theta}$, we find 
that fitting $\theta$ as a function of $\alpha$ produces results that agree well with the exact value.

Finally, following a similar procedure for the corner Renyi entropies of the $2+1$ TFIM
at its quantum critical point, we extrapolate existing NLCE data.  Using the optimal $\theta$ found for free bosons,
but including a non-trivial scaling dimension for the energy operator, we find that the values for $a_{\alpha}(\pi/2)$ 
are much closer to the free boson values than concluded from previous analyses. 

Improving numerical accuracy with such extrapolations is crucially important for the interpretation of many 
 lattice calculations of Renyi entropies.  For example, recent calculations on several models 
tuned to the $O(N)$ Wilson-Fisher fixed-point have revealed 
that $a_{\alpha}(\pi/2)$ is a universal function of $\alpha$ that scales to within numerical accuracy
as \mbox{$a_{\alpha}(\pi/2) \sim N c_{\alpha}(\pi/2)$}, where $N$  is the number of field components, 
and $c_{\alpha}$ is some universal function that appears to be the same for all $O(N)$ models studied to date. 
Intriguingly, the central charge $C_T$ also grows linearly with $N$ up to small corrections \cite{Kos:2014,Bueno:2015}.
Our results indicate that, when unusual corrections to scaling are taken into account, $a_{\alpha}(\pi/2)$ for $N=1$ 
can be very close numerically to the value for free bosons, however for other $O(N)$ values, the conclusion that
\mbox{$a_{\alpha}(\pi/2) \sim N c_{\alpha}(\pi/2)$} remains well justified.
Thus, we confirm that the universal subleading corner coefficient is sensitive to number of degrees of freedom in the low-energy field theory 
describing a strongly-interacting quantum system at its critical point.

\subsection*{Acknowledgments} 
We would like to acknowledge crucial discussions with J.~Cardy, P.~Fendley, R.~Myers, F.~Pollmann, and W.~Witczak-Krempa.
This research was supported in part by the National Science Foundation Grant No. NSF PHY11-25915 (RRPS, RGM), NSF DMR-1306048 (RRPS) and Grant No. 1006549.
The simulations were performed on the computing facilities of SHARCNET and on the Perimeter Institute HPC.  Support was provided by NSERC, the Canada Research Chair program, the Ontario Ministry of Research and Innovation, the John Templeton Foundation, and the Perimeter Institute (PI) for Theoretical Physics. Research at PI is supported by the Government of Canada through Industry Canada and by the Province of Ontario through the Ministry of Economic Development \& Innovation.  

\appendix

\section{Reduced density matrices and entropies for free theories}\label{app:freeboson}

We recall in this appendix some general results regarding reduced density matrices for free bosonic and fermionic systems. The techniques are well-known; we refer to Ref.~\cite{EislerPeschelreview} for a review. We first present the method to compute entanglement for free bosons, relevant to Sec.~\ref{sec:fsft}, before focusing on Renyi correlations (relevant to Sec.~\ref{sec:freefermions}).
\subsection{Entropies for the free lattice scalar field}

It has been long known how to evaluate the entanglement entropy for a system of coupled oscillators, see e.g. Refs.~\cite{Bombelli1986,Srednicki1993}. Here we follow the method put forward by Peschel \cite{Peschel}, that was used by Casini and Huerta \cite{Casini:2007} to compute exactly the corner contribution mentioned in the text. Numerically, the technique may also be used to compute the entropy on very large clusters.

Consider a finite two-dimensional square lattice, such that a free scalar field $\phi_i$ and its conjugate momentum $\pi_i$ exist at each lattice point.
A non-interacting system is described by following lattice Hamiltonian, 
\begin{align}\label{eq:bosonhamiltonian}
H = \frac{1}{2} \sum _{x,y =1,1}^{N_{x},N_{y}} \Big[ \pi_{x,y}^{2} &+  (\phi_{x+1,y} - \phi_{x,y})^{2} + 
 (\phi_{x,y+1} - \phi_{x,y})^{2}\nonumber\\ +& \;m^{2} \phi _{x,y} ^{2}\Big],
\end{align} 
where $N_x$ and $N_y$ are the linear dimensions of the lattice. The Hamiltonian can be rewritten in terms of bosonic operators, but we refrain from doing so for now. We are mainly interested in the massless case $m=0$, but keep general $m$ in all the formulas. 

The total number of sites is $N_x N_y$. The Hamiltonian may be rewritten as
\begin{equation}
 H=\frac{1}{2} \sum_{i=1}^{N_x N_y} \pi_i^2 + \sum_{i,j=1}^N \phi_i M_{ij} \phi_j
\end{equation}
The matrix $M$ is, up to a diagonal term, nothing but the discrete laplacian. The ground state correlations are given
\begin{align}\label{eq:x}
 X_{ij}&=\braket{\phi_i \phi_j}=\frac{1}{2}(M^{-1/2})_{ij},\\\label{eq:p}
 P_{ij} &=\braket{\pi_i \pi_j}=\frac{1}{2}(M^{1/2})_{ij},
\end{align}
and may be obtained explicitly in many geometries, numerically in others. 

Let us now cut our system in two parts $A$ and $B$, and focus only on the degrees of freedom inside subsystem $A$. Since $\braket{\hat{O}}={\rm Tr}\, (\rho_A \hat{O})$ for any observable $\hat{O}$ in subsystem $A$, the reduced density matrix
 itself is the exponential of a quadratic boson form
\begin{equation}\label{eq:rhoabosons}
\rho_{A} = \frac{1}{Z} e^{- \sum_{q} \epsilon_{q} b_{q}^{\dagger} b_{q} },
\end{equation}
where the operators $b_q,b_q^\dag$ are linear combinations of the original fields $\phi_i, \pi_i$. They also satisfy the canonical commutation relations $[b_q,b_{q'}^\dag]=\delta_{qq'}$. Here 
$Z$ 
is a normalization constant that ensures ${\rm Tr}(\rho_A)=1$.
Such an ansatz has by definition the Wick factorization property build in. The next step is then to adjust the energies $\epsilon_l$ and the operators $b_l^\dag$ so as to reproduce the ground-state correlations, which can always be done.
After some algebra, we find that the single particle energies $\epsilon_l$ satisfy
\begin{equation}
(1/2) {\rm coth} (\epsilon_q/2) = \nu_q,
\end{equation}
where the $\nu_q$ are the positive eigenvalues of the matrix $C_A=\sqrt{X_A P_A}$. Here $X_A$ (resp. $P_A$) is the matrix $X$ (resp. $P$) restricted to subsystem $A$. We note it $C_A$, to emphasize the fact that it depends on the choice of subsystem. The Renyi entropy then follows from Eq.~(\ref{eq:rhoabosons})
\begin{equation}\label{eq:entropyfreeboson}
 S_\alpha=\frac{1}{\alpha -1} {\rm Tr} \log \left[\left(C_A + 1/2\right)^\alpha  - \left(C_A - 1/2\right)^\alpha\right].
\end{equation}
This is of course a huge simplification, because what is only needed is the diagonalization of the matrix $C_A$, whose size is given by the number of sites in $A$. This last step can be done using standard linear algebra routines.

For the geometry we choose the full system to be an open-boundary rectangle of size $N_x\times N_y$. The correlators come from diagonalizing the discrete laplacian on the rectangle, and using Eqs.~(\ref{eq:x}, \ref{eq:p}). We obtain

\begin{eqnarray}
\langle \phi_{x,y}\phi_{x',y'} \rangle & = &\frac{2}{(N_x+1)(N_y+1)}\sum_{q_x,q_y}\sin\left(q_{x} x\right) \sin\left(q_{x} x'\right) \nonumber\\
& &  \times \sin\left(q_{y} y\right) \sin\left(q_{y} y'\right) \frac{1}{\omega(q_x,q_y)} \label{OBCphi} \\
\langle \pi_{x,y}\pi_{x',y'} \rangle & = & \frac{2}{(N_x+1)(N_y+1)}\sum_{q_x,q_y} \sin\left(q_{x}x\right) \sin\left(q_{x} x'\right) \nonumber\\
& & \times \sin\left(q_{y} y\right) \sin\left(q_{y} y'\right) \times \omega(q_x,q_y) \label{OBCpi}
\end{eqnarray} 
where the quasi-momenta $k_x,k_y$ are quantized as
\begin{eqnarray}
 q_x&=&\frac{n_x\pi}{N_x+1}\quad,\quad n_x=1,2,\ldots,N_x,\\
  q_y&=&\frac{n_y\pi}{N_y+1}\quad,\quad n_y=1,2,\ldots,N_y.
\end{eqnarray}
and
\begin{equation}
 \omega(q_x,q_y)=\sqrt{4\sin^2 (q_x/2)+4\sin^2(q_y/2)+m^2}.
\end{equation}
Combined with the NLCE procedure described in Appendix~\ref{AppNLCE}, this produces the data shown in Sec.~\ref{sec:fsft}.
 We note that the choice of boundary conditions is not completely innocent. For example, if we were to choose a periodic-boundary
 torus instead of an open-boundary rectangle, the correlators would be given by similar but translation invariant formulas
\begin{align}
\langle \phi_{0,0}\phi_{x,y} \rangle & = \frac{1}{2 N_xN_y}\sum_{k_x,k_y}\frac{\cos\left(k_{x} x\right) \cos\left(k_{y} y\right)}{\omega(k_x,k_y)}, \label{PBCphi} \\
\langle \pi_{0,0}\pi_{x,y} \rangle & =  \frac{1}{2 N_xN_y}\sum_{k_x,k_y} \cos\left(k_{x} x\right)\cos\left(k_{y} y\right) \omega(k_x,k_y), \label{PBCpi}
\end{align} 
with
\begin{equation}
  \omega(k_x,k_y)=\sqrt{4\sin^2 (k_x/2)+4\sin^2(k_y/2)+m^2}.
\end{equation}
and the momenta quantized as
\begin{eqnarray}
 k_x&=&\frac{2n_x\pi}{N_x}\quad,\quad n_x=0,1,\ldots,N_x-1,\\
  k_y&=&\frac{2n_y\pi}{N_y}\quad,\quad n_y=0,1,\ldots,N_y-1.
\end{eqnarray}
Unfortunately, the correlator (\ref{PBCphi}) is divergent in the massless case $m=0$, due to the presence of a zero mode $(k_x,k_y)=(0,0)$.
Thus, $S_{\alpha}$ also diverges, as shown previously in Ref.~\cite{Max_Tarun}, significantly complicating the calculation of Renyi entropies
on finite-size tori. To remove this divergence, a possible way is keep a very small mass term; this results in an additional term
proportional to $\log(mL)$ in the entropy, where $L$ is the length of the boundary. This $m$ may then be systematically
decreased and the results extrapolated to $m \rightarrow 0$. 

This complication is not a fundamental one; however, it motivates us to only consider clusters with open boundary conditions
in the procedure of Appendix \ref{AppNLCE}, which do not have zero modes.

\subsection{Renyi correlations}
\label{app:renyicorr}
We consider the following Hamiltonian
\begin{equation}\label{eq:freefermionHamiltonian}
 H=\sum_{i,j=1}^{L} \left(U_{ij} c_{i}^\dag c_j +V_{ij} c_i^\dag c_{j}^\dag +h.c\right).
\end{equation}
Such a quadratic fermion form can always be diagonalized by a Bogoliubov transformation. Namely, there is a set of fermion operators
\begin{equation}
 d_k^\dag=\sum_{j}\alpha_{kj}c_j^\dag+\beta_{kj}c_j
\end{equation}
which obey the canonical commutation relations $\{d_k^\dag,d_{k'}\}=\delta_{kk'}$, and which diagonalize $H$:
\begin{equation}
 H=\sum_k \epsilon_k d_{k}^\dag d_k
\end{equation}
The coefficients $\alpha_{kj},\beta_{kj}$ and the single particle energies $\epsilon_k$ may be e.g. obtained by diagonalizing the following $2L\times 2L$ Bogoliubov matrix
\begin{equation}
 M=\left(
 \begin{array}{cc}
  U&V\\
  V^*&1-U^t
 \end{array}
 \right)
\end{equation}
where $U^t$ (resp. $V^*$) is the transpose of $U$ (resp. conjugate transpose of $V$). Using this method, computing for example the ground-state correlations is a straightforward task.

We now focus on a subsystem, which we choose to be the first $\ell$ sites for simplicity. For later convenience we introduce the following $2\ell\times 2\ell$ correlation matrix
 \begin{equation}\label{eq:korrmatrix}
 K=\left(\begin{array}{cc}
          C&D\\D^*&1-C^t
         \end{array}
\right).
\end{equation}
$C$ and $D$ encode the two types of correlators:
\begin{eqnarray}
 C&=&(\braket{c_i^\dag c_j})_{1\leq i,j\leq \ell},\\
 D&=&(\braket{c_i^\dag c^\dag_j})_{1\leq i,j\leq \ell}.
\end{eqnarray}
Now, any correlation in subsystem $A$ may be recovered from the reduced density matrix, 
\begin{equation}\label{eq:someO}
 \braket{O}={\rm Tr}\, (\rho_A O).
\end{equation}
Because of Wick's theorem, the reduced density matrix itself can be written as
\begin{equation}
\rho_A=\frac{e^{-H_A}}{{\rm Tr}\,e^{-H_A}},
\end{equation}
where $H_A$ is a quadratic fermion form for a system of size $\ell$, similar to (\ref{eq:freefermionHamiltonian}). We name $H_A$ the entanglement Hamiltonian, as it is different from the original Hamiltonian $H$. The crucial point is that the Bogoliubov matrix corresponding to $H_A$ and the subsystem correlation matrix are diagonalized by the same transformation \cite{Peschel}. More precisely, it is possible to find a unitary matrix $U$ which diagonalizes $K$
\begin{equation}\label{eq:kdiag}
 K=U^{*} \left(\begin{array}{cc}
                  {\rm diag}_q\left(\lambda_q\right)&0\\0&{\rm diag}_q\left(1-\lambda_q\right)
                 \end{array}
\right)U,
\end{equation}
and such that 
\begin{equation}
 H_A=\sum_{q=1}^{\ell} \varepsilon_q f_q^\dag f_q\quad,
\end{equation}
with
\begin{equation}\label{eq:transf}
 (f_1^\dag,\ldots ,f_\ell^\dag,f_1,\ldots,f_\ell)^t=U (c_1^\dag,\ldots ,c_\ell^\dag,c_1,\ldots,c_\ell)^t.
\end{equation}
The single particle eigenenergies $\varepsilon_q$ of the entanglement Hamiltonian may then be obtained from the $\lambda_q$ by requiring consistency with (\ref{eq:someO}). We obtain
\begin{equation}\label{eq:relation}
 \lambda_q=\frac{1}{1+e^{\varepsilon_q}}.
\end{equation}
This relation, together with (\ref{eq:transf}), uniquely determines the entanglement Hamiltonian from the eigenvalues and the eigenvectors of the correlation matrix $K$. 

With this correspondence at hand, computing Renyi correlations in subsystem $A$ becomes straightforward. Indeed the new (powered up) reduced density $\rho_A^{(\alpha)}$ is
\begin{equation}
 \rho_A^{(\alpha)}=\frac{(\rho_A)^\alpha}{{\rm Tr}\left[(\rho_A)^\alpha\right]}=\frac{e^{-\alpha H_A}}{{\rm Tr}\left[ e^{-\alpha H_A}\right]}.
\end{equation}
Therefore, the new entanglement Hamiltonian is simply $\alpha H_A$. We now wish to obtain the new correlation matrix 
\begin{equation}
 K_\alpha=\left(\begin{array}{cc}
          C_\alpha&D_\alpha\\D_\alpha^*&1-C^t_\alpha
         \end{array}
\right),
\end{equation}
with
\begin{eqnarray}
 C_\alpha&=&(\braket{c_i^\dag c_j}_\alpha)_{1\leq i,j\leq \ell},\\
 D_\alpha&=&(\braket{c_i^\dag c^\dag_j}_\alpha)_{1\leq i,j\leq \ell},
\end{eqnarray}
corresponding to the Renyi correlators
\begin{equation}
 \braket{O}_\alpha={\rm Tr}\left[\rho_A^{(\alpha)}O\right].
\end{equation}
To do so it is sufficient to reverse the logic leading to the determination of $H_A$. From (\ref{eq:kdiag}) and (\ref{eq:relation}), we have
\begin{equation}
 K=U^{*} \left(\begin{array}{cc}
                  {\rm diag}_q\left(\frac{1}{1+e^{\varepsilon_q}}\right)&0\\0&{\rm diag}_q\left(\frac{1}{1+e^{-\varepsilon_q}}\right)
                 \end{array}
\right)U
\end{equation}
so the new correlation matrix $K_\alpha$ may be obtained by simply multiplying all single particle energies $\varepsilon_q$ by $\alpha$. We obtain
\begin{equation}
 K_\alpha=U^{*} \left(\begin{array}{cc}
                  {\rm diag}_q\left(\frac{1}{1+e^{\alpha\varepsilon_q}}\right)&0\\0&{\rm diag}_q\left(\frac{1}{1+e^{-\alpha\varepsilon_q}}\right)
                 \end{array}
\right)U.
\end{equation}
In terms of the initial correlation matrix $K$, the previous relation reads
\begin{equation}\label{eq:mainresult}
 K_\alpha=\left[1+(K^{-1}-1)^\alpha\right]^{-1}.
\end{equation}
The result is true for any eigenstate and also at finite temperature, the only requirement being a free fermion Hamiltonian. In fact, (\ref{eq:mainresult}) even holds for quadratic boson Hamiltonians, provided the correlation matrix be redefined as
\begin{equation}
 K=\left(\begin{array}{cc}
          C^t&D\\D^*&1+C^t
         \end{array}
\right).
\end{equation}
instead of (\ref{eq:korrmatrix}). 
This can be shown by repeating the steps described above with bosonic operators instead of fermions.

\section{Series analysis} \label{AppA}

Series are analyzed by the method of Differential Approximants \cite{fisher,baker}.
These are generalizations of the better
known d-log Pade method and allow for a power-law singularity. 
The function of interest $f(x)$ is
represented by a solution to a first order ordinary differential equation:
\begin{equation}
P_1^M(x) {df(x)\over dx} + P_2^N(x) f(x) = P_3^J(x)
\end{equation}
Here, $P_1^M(x)$, $P_2^N(x)$ and $P_3^J(x)$ are polynomials of order $M$, $N$ and $J$ respectively
that are determined by demanding that the series
expansion for the function $f(x)$ match to some order. It is easy to show that these differential
equations have power-law singularities at critical points given by the roots of the polynomial $P_1^M(x)$
with exponents related to the residue of $P_1^M(x)$ and the value of $P_2^N(x)$ at that critical point.
The less important $P_3^J(x)$ term allows for a background function.
Since the critical point of Eq.~(\ref{eq:ictf_hamiltonian}) is known to be at $J/h=1$, we further bias the analysis by demanding that
\begin{equation}
P_1^M(x=1)=0,
\end{equation}
thus ensuring that critical point is at $x_c=1$.

The calculation proceeds by first choosing orders for the polynomials $M$, $N$ and $J$. We fix $J=0$
that just allows for a constant background term, and keep $M$ to at least $2$.
The sum of the orders ($M+N+J$) can not exceed the number of terms (more correctly number of terms minus one after the biasing)
that one knows in the series expansion. 
In general, different choice of the order of the three polynomials $M$, $N$ and $J$ will give different values
for critical exponents.

\section{The Numerical Linked Cluster Expansion} \label{AppNLCE}

The procedure of the Appendix \ref{app:freeboson} allows one to calculate the entanglement entropy for free bosons on clusters of arbitrary shape and size.
Thus, it may be used as a ``cluster solver'' for a NLCE procedure, similar to previous studies where Lanczos or DMRG were used as cluster solvers for interacting models \cite{Kallin:2013,Kallin:2014,Miles:2014}.

We refer the reader to the relevant literature \cite{Rigol:2006, Rigol:2007_1,Rigol:2007_2,Tang:2013}. 
The NLCE method is based on the fact that an extensive property $P$ of a lattice model can be expressed as a 
sum over contributions from all distinct clusters which are embeddable in the lattice.    Then, this property per site is,
\begin{equation}
P/N = \sum_c W(c),
\end{equation}
where $W(c)$ is the ``weight'', or the {\em unique} contribution to $P$ that comes from a cluster $c$ \footnote{Note, unique embedding of the same cluster shape must be accounted for by the sum.}.  It is defined recursively,
\begin{equation}
W(c) = P(c) - \sum_{s \in c} W(s),
\end{equation}
where $s$ is any subcluster of $c$.  As dictated by the inclusion-exclusion principle, only connected clusters give non-zero weights to the sum.
The NLCE procedure builds up the value of $P/N$ starting from the smallest cluster $c=1$ and ending with some maximal cluster size.
For this paper, we employ only rectangular clusters \cite{Kallin:2013}, which makes the calculation of cluster embeddings $s \in c$ trivial; then, the computational
bottleneck is shifted to the cluster solver.

For interacting models, the computational cost of calculating the given property $P$ on the cluster is extremely expensive.  However, 
as described in the last section, a very efficient numerical method can be devised for calculating the Renyi entropies for a free bosons
on the square lattice using Equations (\ref{OBCphi},\ref{OBCpi}).  This allows us to perform the NLCE to extremely high orders.

As with any cluster solver on the square lattice, the NLCE is able to isolate the entanglement due to a $\pi/2$ corner 
by subtracting off the contributions to $S_{\alpha}$ from the linear portions of the boundary (the {\em area-law} term in Eq.~(\ref{EEscale})).
As described in Ref.~\cite{Kallin:2013} the property $P$ then becomes the isolated contribution for a single corner, given by Eq.~(\ref{eq:Corner}).  
This property is made
extensive by adding the values for all possible translations of the corner around the cluster $c$.

Finally, for studies of critical systems such as the free boson, the definition of a length scale is important.  In this paper, we are interested in 
benchmarking the success of the NLCE in converging the exact results for $a_{\alpha}(\pi/2)$ in the limit where this 
length scale approaches infinity.  A second advantage of using rectangular clusters is the easy definition of this length scale
in relation to the cluster order $O$.
In past studies, several definitions have been used, including the arithmetic and geometric means of the rectangle linear dimensions \cite{Kallin:2013,Kallin:2014,Miles:2014,Kallin_thesis}.
In the main text, we define the cluster length scale $\ell$ as the maximum of the two linear dimensions. This definition,
which also defines the expansion order $O$, is simple to use
and is motivated by the intuition that the role of the smaller clusters in the NLCE sum is to cancel boundary effects. Thus long and narrow
clusters of dimension $N_x, N_y$ should only be included at the same order as $N \times N$ square clusters with $N = \max(N_x,N_y)$;
otherwise these narrow clusters are not the boundary of any more isotropic cluster appearing to this order in the sum.

\bibliographystyle{apsrev}
\bibliography{fscorn}

\end{document}